\documentclass[runningheads]{llncs}
\usepackage[T1]{fontenc}
\usepackage{graphicx}
\usepackage[table]{xcolor}
\usepackage{hyperref}

\usepackage{multirow}
\usepackage{tabu}
\usepackage{bbm}
\usepackage{diagbox}
\usepackage{comment}
\usepackage{array}
\usepackage{amstext}
\usepackage{makecell}
\usepackage{tablefootnote}
\usepackage{booktabs} 
\usepackage{color}
\usepackage{amsmath}
\usepackage{wrapfig}
\usepackage{amssymb}
\usepackage{pifont}
\usepackage[backend=biber]{biblatex}
\newcommand{\ourmodel}{CAFusion}

\begin{document}
\title{\ourmodel{}: Controllable Anatomical Synthesis of Perirectal Lymph Nodes via SDF-guided Diffusion}
\titlerunning{\ourmodel{}: Controllable Anatomical Synthesis}

\authorrunning{W. Guo et al.}
\author{Weidong Guo\inst{1,2} \and
Hantao Zhang\inst{1,2} \and
Shouhong Wan\inst{1,2}\thanks{Corresponding author: \email{wansh@ustc.edu.cn}} \and
Bingbing Zou\inst{3,2,4} \and
Wanqin Wang\inst{3,2,4} \and
Chenyang Qiu\inst{3,2,4} \and
Peiquan Jin\inst{1}}
\institute{
School of Computer Science and Technology, University of Science and Technology of China, Hefei, China
\and
Institute of Artificial Intelligence, Hefei Comprehensive National Science Center, Hefei, China
\and
Department of General Surgery, The First Affiliated Hospital of Anhui Medical University, Hefei, China
\and
Anhui Medical University, Hefei, China
}
\maketitle
\begin{abstract}
Lesion synthesis methods have made significant progress in generating large-scale synthetic datasets. However, existing approaches predominantly focus on texture synthesis and often fail to accurately model masks for anatomically complex lesions. Additionally, these methods typically lack precise control over the synthesis process. For example, perirectal lymph nodes, which range in diameter from 1 mm to 10 mm, exhibit irregular and intricate contours that are challenging for current techniques to replicate faithfully.
To address these limitations, we introduce \ourmodel{}, a novel approach for synthesizing perirectal lymph nodes. By leveraging Signed Distance Functions (SDF), \ourmodel{} generates highly realistic 3D anatomical structures. Furthermore, it offers flexible control over both anatomical and textural features by decoupling the generation of morphological attributes—such as shape, size, and position—from textural characteristics, including signal intensity.
Experimental results demonstrate that our synthetic data substantially improve segmentation performance, achieving a 6.45\% increase in the Dice coefficient. In the visual Turing test, experienced radiologists found it challenging to distinguish between synthetic and real lesions, highlighting the high degree of realism and anatomical accuracy achieved by our approach. These findings validate the effectiveness of our method in generating high-quality synthetic lesions for advancing medical image processing applications. Our code will be available at \url{https://anonymous.4open.science/r/CAFusion-C7D3}.
\end{abstract}

\section{Introduction}
    The development of robust foundational models in medical imaging hinges on large, accurately annotated datasets. However, the high cost and labor-intensive nature of manual annotations severely restrict dataset availability\cite{aljabri2022towards,li2018large}. Compounding this challenge, inherent biases in patient data collection across healthcare systems frequently yield lesion datasets with long-tailed distributions\cite{perets2024inherent}, creating significant bottlenecks for model training\cite{zhang2023deep,wu2024medical}.

Medical lesion synthesis has emerged as a promising solution, enabling scalable generation of synthetic lesions to augment limited datasets\cite{Savage2023SyntheticDC}. State-of-the-art methods typically follow a two-stage pipeline: (1) mask synthesis to define lesion morphology and (2) texture synthesis guided by the mask to generate realistic lesions. While these approaches have proven effective across diverse applications---including tumors\cite{jin2021free,hu2023label,chen2024towards}, lung nodules\cite{han2019synthesizing,wang2021realistic}, colon polyps\cite{shin2018abnormal,sharma2024controlpolypnet}, and diabetic retinopathy\cite{zhou2019high,abbood2022dr}---a critical gap persists. Most efforts prioritize texture realism, leaving mask synthesis comparatively underexplored despite its foundational role in ensuring lesion diversity and anatomical plausibility.

Current mask synthesis strategies suffer from two key limitations. First, methods that directly extract masks from real data\cite{yao2021label} inherently restrict diversity due to dataset biases. Second, rule-based designs\cite{hu2023label} lack the flexibility to capture complex morphological features (e.g., irregular edges, fine structural variations), sacrificing authenticity and generalizability. Suboptimal masks not only degrade synthetic lesion fidelity but also propagate errors to downstream tasks reliant on precise annotations, such as lesion segmentation\cite{zhu2004class,ju2022improving}. These shortcomings underscore the urgent need for controllable, adaptive mask synthesis methods capable of generating diverse and anatomically realistic 3D lesion masks.

To address this gap, we propose a novel framework for controllable 3D synthesis of perirectal lymph node morphology, a clinically relevant yet morphologically complex target. perirectal lymph nodes in CT imaging exhibit significant diversity in shape (highly irregular edges) and size (1--10 mm diameter)\cite{kim2004high}, necessitating synthesis methods that balance precision and adaptability. Existing handcrafted approaches, constrained by rigid rules, fail to capture these intricate features, limiting their utility in both research and clinical applications. Our work leverages signed distance functions (SDF)\cite{curless1996volumetric,park2019deepsdf} and diffusion models\cite{ho2020denoising,song2020denoising} to enable training-free, fine-grained control over 3D morphology. By integrating morphological priors as explicit constraints, we achieve flexibility in synthesizing realistic and diverse lesions while maintaining computational efficiency.

The key contributions of this paper are as follows:

\begin{itemize}
    \item \textbf{SDF-based Method for Controllable 3D Synthesis.} 
    We present an SDF-guided approach for synthesizing perirectal lymph node morphology. Unlike voxel-based representations, which often face discontinuity and instability, our method leverages the continuous properties of signed distance functions to model lymph node shapes. This enables the precise generation of complex edges, irregular surfaces, and fine-grained morphological details, providing a robust foundation for controllable synthesis.

    \item \textbf{Controllable Synthesis Framework.} 
    Building on SDF-based modeling, we propose a novel synthesis framework that combines loss-guidance \cite{song2023loss} and repaint \cite{lugmayr2022repaint}. Loss-guidance ensures that lymph node shape, size, and signal intensity align with control parameters, maintaining consistency with desired anatomical characteristics. Repaint seamlessly integrates the synthesized lymph nodes into CT scan backgrounds, ensuring visual and structural coherence with surrounding tissues.

    \item \textbf{Enhanced Diversity and Segmentation Performance.} 
    Our method generates lymph nodes with diverse variations in location, size, shape, and signal intensity. Experiments demonstrate that the synthesized lymph nodes are visually indistinguishable from real ones, as confirmed by radiologists. Furthermore, the inclusion of diverse synthetic data significantly enhances segmentation performance in downstream tasks.
\end{itemize}
% !TEX root = ../top.tex
% !TEX spellcheck = en-US

\begin{figure}[tb]
        \centering
	\includegraphics[width=\linewidth]{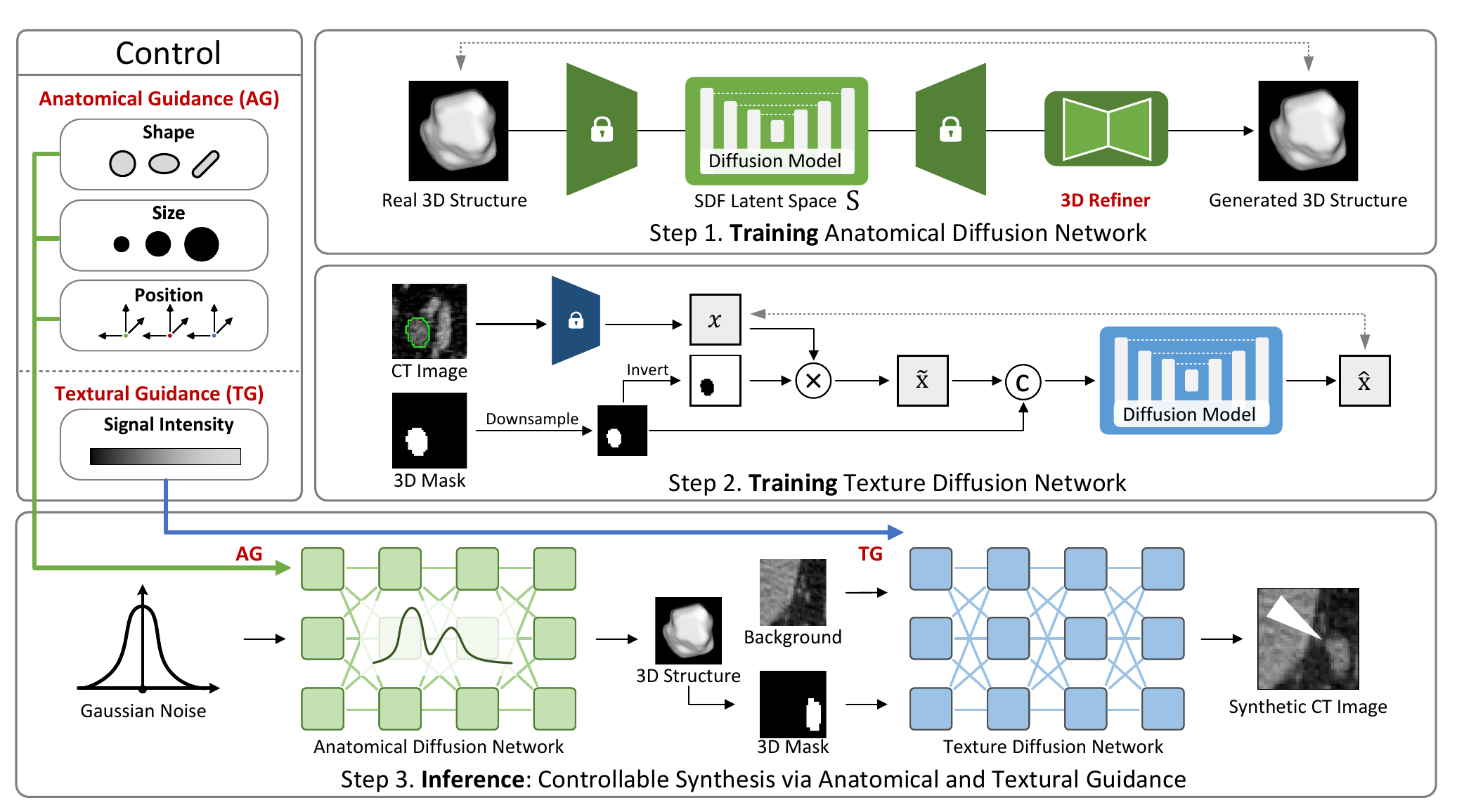}
	\caption{\textbf{Illustration of Controllable Perirectal Lymph Node Synthesis.} Our method consists of three key steps. First, we train a network to generate the anatomical structure of perirectal lymph nodes, establishing a morphology prior. Second, we train an image inpainting network to construct a texture prior. Finally, we perform controllable synthesis by incorporating anatomical guidance during mask generation and textural guidance during texture generation. }
	\label{fig:framework}
\end{figure}
\section{Method}
\subsection{Background and Overview}

Existing methods for 3D lesion synthesis \cite{jin2021free,hu2023label,chen2024towards} focus primarily on generating realistic textures but struggle to accurately control 3D shapes. Precise control over lesion morphology is crucial for producing anatomically realistic and diverse samples.

To address this challenge, we propose \ourmodel{}, a framework for controllable synthesis of perirectal lymph nodes. As shown in Figure~\ref{fig:framework}, our method integrates \textbf{anatomical guidance (AG)} for precise control of lymph node shapes. Additionally, we introduce \textbf{textural guidance (TG)} to ensure the synthesized lymph nodes exhibit realistic signal intensities. This dual-guidance framework allows us to generate lymph nodes with variations in cross-sectional shape, surface irregularities, location, and signal intensity, achieving both realism and controllability.

\subsection{\ourmodel{} Framework}

\paragraph{\textbf{Anatomical Guidance (AG).}}
The Signed Distance Function (SDF) \cite{curless1996volumetric} provides a continuous representation of 3D anatomical structures by encoding the distance from any point to the surface. Leveraging this property, we embed the SDF in a latent space \( \mathbf{s} \in \mathcal{S} \), where both sampling and guidance are performed to ensure anatomical accuracy.

To enforce anatomical constraints during synthesis, we define an anatomical guidance loss \(\mathcal{L}_{\text{AG}}\), which incorporates both global and local constraints:
\begin{equation}
\mathcal{L}_{\text{AG}} = \lambda_1 \| \hat{\mathbf{s}}_0 - \mathbf{s}_{\text{target}} \|^2 + \lambda_2 \| \text{CI}(\hat{\mathbf{s}}_0) - \text{CI}(\mathbf{s}_{\text{target}}) \|^2,
\end{equation}
where the first term enforces global shape fidelity, and the second term refines surface details. The Curvature Index (CI), quantifying surface irregularities, is defined as:
\begin{equation}
\text{CI} = \|\nabla \hat{\mathbf{n}}\|_F = \sqrt{\sum_{i,j} \left( \frac{\partial \hat{n}_i}{\partial x_j} \right)^2},
\end{equation}
with \(\hat{\mathbf{n}}\) as the normalized surface normal. The weights \(\lambda_1\) and \(\lambda_2\) balance global shape preservation and local detail refinement.

During inference, \(\mathcal{L}_{\text{AG}}\) is computed at each time step \( t \) of the sampling process. The latent variable at step \( 0 \), \(\hat{\mathbf{s}}_0\), is obtained via denoising from the latent variable at step \( t \). To improve stability, Monte Carlo sampling is used, averaging the loss over multiple independent denoising trajectories.

The anatomical guidance is applied by calculating the gradient of the loss to adjust the latent variable \( \mathbf{s}_t \). This produces a guidance-adjusted variable \(\tilde{\mathbf{s}}_t\):
\begin{equation}
\tilde{\mathbf{s}}_t = \mathbf{s}_t - \eta_t \nabla_{\mathbf{s}_t} \mathcal{L}_{\text{AG}}.
\end{equation}
The adjusted variable \(\tilde{\mathbf{s}}_t\) is then used to predict the mean \(\boldsymbol{\mu}_\theta(\tilde{\mathbf{s}}_t, t)\) and variance \(\boldsymbol{\sigma}_\theta^2(\tilde{\mathbf{s}}_t, t)\), defining the Gaussian distribution for sampling \( \mathbf{s}_{t-1} \):
\begin{equation}
\mathbf{s}_{t-1} = \mathcal{N}(\boldsymbol{\mu}_\theta(\tilde{\mathbf{s}}_t, t), \boldsymbol{\sigma}_\theta^2(\tilde{\mathbf{s}}_t, t)).
\end{equation}

\paragraph{\textbf{Textural Guidance (TG).}}
To ensure robust and controllable texture synthesis, TG applies a signal intensity (SI) constraint, aligning the generated textures with a target intensity distribution. The textural guidance loss, defined as the MSE between the predicted image \(\hat{\mathbf{x}}_0\) and the target intensity \(\text{SI}_{\text{target}}\), is:
\begin{equation}
\mathcal{L}_{\text{TG}} = \| \text{SI}(\hat{\mathbf{x}}_0) - \text{SI}_{\text{target}} \|^2.
\end{equation}
Following a similar process to AG, the latent variable \(\mathbf{x}_t\) is adjusted using the gradient \(-\gamma_t \nabla_{\mathbf{x}_t} \mathcal{L}_{\text{TG}}\), producing a guidance-adjusted variable \(\tilde{\mathbf{x}}_t\) for sampling. This ensures the synthesized textures adhere to the desired intensity distribution while maintaining stability.

\paragraph{\textbf{Masked Repaint with Background Preservation.}}
To seamlessly integrate synthesized textures into the anatomical context, \ourmodel{} employs a masked repaint strategy. The final output at timestep \( t-1 \) is:
\begin{equation}
\mathbf{x}_{t-1}^\text{repaint} = \mathbf{x}_{t-1} \odot \mathbf{M} + \mathbf{x}_{t-1}^\text{bg} \odot (1 - \mathbf{M}),
\end{equation}
where \( \mathbf{M} \) is the lesion mask, and \( \mathbf{x}_{t-1}^\text{bg} \) represents the background features obtained by adding noise up to step \( t-1 \). This ensures that the background remains anatomically consistent, while the foreground lesion region is synthesized with guided structural and textural features.
\section{Experiments}

\subsection{Dataset}
We extend the work of CARE\cite{zhang2023care} by incorporating perirectal lymph node annotations. The perirectal lymph node dataset comprises CT volumes from 100 patients, with a total of 1,137 voxel-level annotations of perirectal lymph nodes. The dataset is split into a training set of 80 cases and a test set of 20 cases. Lymph nodes were synthesized using the training set, and both the synthesis network and the segmentation network were evaluated on the test set.

\subsection{Improving Segmentation Performance}
\setlength{\tabcolsep}{0.5pt}
\begin{table}[tb]
    \caption{\textbf{Segmentation Performance.} This table compares segmentation performance metrics for various methods trained with different synthetic data strategies. \textbf{Baseline}: training on the original dataset without synthetic augmentation. \textbf{Copy-Paste}: inserting real lymph node patches into anatomically appropriate locations. \textbf{Cond-Diffusion} \cite{chen2024towards}: generating synthetic lymph nodes conditioned on lesion masks and background images. \textbf{Repaint} \cite{lugmayr2022repaint}: reconstructing lesion areas by blending lesion foregrounds with diffused backgrounds. \textbf{\ourmodel{}} (proposed): synthesizing both lymph node masks and corresponding CT scans. All methods augment the original dataset with 300 synthetic CT scans, containing approximately 1,400 synthetic lymph nodes. \colorbox{green!15}{\textbf{Bold}} values denote the best performance for each metric, while \colorbox{red!25}{red} highlights indicate significant improvements over the baseline.}
	\label{tab:segmentation_performance}
	\centering
    \begin{tabular*}{\hsize}{@{}@{\extracolsep{\fill}}lccccccccc@{}}
    	\toprule
    	\multirow{2}{*}{\textbf{Method}} & \multicolumn{3}{c}{\textbf{nnUNet\cite{isensee2021nnu}}} & \multicolumn{3}{c}{\textbf{UNet\cite{ronneberger2015u}}} & \multicolumn{3}{c}{\textbf{SwinUNETR\cite{he2023swinunetr}}} \\
    	\cmidrule(lr){2-4} \cmidrule(lr){5-7} \cmidrule(lr){8-10}
    	& \textbf{Dice} & \textbf{IoU} & \textbf{NSD} & \textbf{Dice} & \textbf{IoU} & \textbf{NSD} & \textbf{Dice} & \textbf{IoU} & \textbf{NSD} \\
    	\midrule
    	Baseline & 54.52 & 37.47 & 49.43 & 52.58 & 35.67 & 44.73 & 53.62 & 36.63 & 45.45 \\
    	\midrule
    	\multicolumn{10}{l}{\textit{Synthetic Data with Real Mask}} \\
    	\midrule
    	Copy-Paste & \cellcolor{red!25}57.46 & \cellcolor{red!25}40.31 & 50.63 & 54.01 & 36.99 & 46.55 & 55.45 & 38.36 & 48.81 \\
    	Cond-Diffusion\cite{chen2024towards}& 57.27 & 40.13 & \cellcolor{red!25}52.26 & \cellcolor{red!25}55.92 & \cellcolor{red!25}38.81 & 47.89 & 53.39 & 36.42 & 45.81 \\
    	Repaint\cite{lugmayr2022repaint} & 56.74 & 39.61 & 50.61 & 53.80 & 36.80 & 47.85 & 54.49 & 37.45 & 47.63 \\
    	\midrule
    	\multicolumn{10}{l}{\textit{Synthetic Data with Generated Mask}} \\
    	\midrule
    	\ourmodel{} (Ours) & 55.13 & 38.05 & 48.81 & 54.02 & 37.01 & 45.43 & \cellcolor{red!25}57.34 & \cellcolor{red!25}40.19 & \cellcolor{red!25}49.76 \\
    	\ourmodel{}+AG (Ours) & 55.48 & 38.39 & 51.37 & 54.90 & 37.83 & \cellcolor{red!25}48.67 & 55.31 & 38.22 & 49.56 \\
    	\ourmodel{}+AG\&TG (Ours) & \cellcolor{green!15}\textbf{60.97} & \cellcolor{green!15}\textbf{43.86} & \cellcolor{green!15}\textbf{52.74} & \cellcolor{green!15}\textbf{56.22} & \cellcolor{green!15}\textbf{39.10} & \cellcolor{red!25}48.67 & \cellcolor{green!15}\textbf{59.30} & \cellcolor{green!15}\textbf{42.14} & \cellcolor{green!15}\textbf{50.63} \\
    	\bottomrule
    \end{tabular*}
\end{table}
\begin{table}[tb]
    \centering
    \caption{\textbf{Ablation Study on Anatomical Synthesis.} This table assesses the contribution of the Refiner (3D refiner module) and AG (anatomical guidance) to the quality of anatomical synthesis. The evaluation metrics include MMD (lower is better), pDSC (lower is better), and COV (higher is better). The best-performing results for each metric are highlighted in \colorbox{green!15}{\textbf{bold}}.}

    \label{tab:ablation_study}
\begin{tabular*}{\linewidth}{@{\extracolsep{\fill}}cc|ccc@{}}
\toprule
\textbf{Refiner} & \textbf{AG} & \textbf{MMD $\downarrow$} & \textbf{pDSC \% $\downarrow$}  & \textbf{COV \% $\uparrow$}\\
\midrule
\ding{55} & \ding{55} & 0.1953 & 69.56 & 43.58 \\
\checkmark & \ding{55} & 0.1796 & 69.20 & 38.07 \\
\checkmark & \checkmark & \cellcolor{green!15}\textbf{0.1219} & \cellcolor{green!15}\textbf{68.63} & \cellcolor{green!15}\textbf{51.83} \\
\bottomrule
\end{tabular*}
\end{table}

We evaluate three perirectal lymph node synthesis paradigms: (1) baseline training with original data, (2) SOTA methods using real masks, and (3) our \textbf{\ourmodel{}} framework generating both masks and CT scans (300 synthetic scans, $\approx$1,400 lymph nodes). As shown in Table~\ref{tab:segmentation_performance}, the baseline \ourmodel{} achieves comparable accuracy to SOTA methods, confirming the anatomical validity of synthetic masks. 

To further improve anatomical diversity, we introduce \textbf{\ourmodel{}+AG}. By incorporating anatomical guidance (AG), the morphological diversity of lymph nodes is significantly enhanced. However, this approach yields only marginal segmentation performance improvements on UNet\cite{ronneberger2015u} and nnUNet\cite{isensee2021nnu}. This reveals a critical limitation: increasing morphological diversity introduces greater challenges in texture generation, as the generative model struggles to maintain texture realism under unseen mask conditions.

To address this issue, we propose \textbf{\ourmodel{}+AG\&TG}, which incorporates both anatomical and textural dual-guidance mechanisms. Textural Guidance (TG) enforces manifold constraints to stabilize texture generation, ensuring consistency with real data distributions even under diverse mask conditions. By constraining mask diversity within anatomically plausible boundaries and guiding texture generation to avoid degradation, the framework ensures both anatomical fidelity and texture realism. This dual-guidance approach effectively decouples the enhancement of morphological diversity from texture degradation, underscoring the critical role of controlled synthesis in producing high-quality and task-effective synthetic data for medical imaging applications.
\subsection{Enhancing Anatomical Synthesis Quality}  

Table~\ref{tab:ablation_study} evaluates the impact of the 3D Refiner module (Refiner) and anatomical guidance (AG) on anatomical structure synthesis. The metrics include\textbf{ Minimum Matching Distance (MMD)} \cite{achlioptas2018learning}, which quantifies the structural discrepancy between generated and real samples; \textbf{Coverage (COV)} \cite{achlioptas2018learning}, which measures the alignment between their data distributions; and \textbf{Pairwise Dice Similarity Coefficient (pDSC)}, which reflects morphological diversity.  

The Refiner reduces reconstruction noise by bridging the domain gap between ShapeNet's pre-trained SDF representation \cite{chang2015shapenet,chou2023diffusion} and medical data, improving synthesis quality. The AG strategy mitigates mode collapse in unconditional generation \cite{bao2022conditional} by sampling morphological features from real data, ensuring better distribution alignment.  

Combining Refiner and AG achieves the best results, enhancing structural fidelity, distribution alignment, and morphological diversity in synthesized anatomical structures.

\subsection{Visual Turing Tests}
\begin{table}[tb]
\centering
\caption{\textbf{Radiologists' performance in the visual Turing test.} Accuracy represents the overall classification correctness, sensitivity quantifies the ability to identify synthetic nodes, and specificity measures the capability to recognize real nodes.}
\label{tab:turing_test}
\begin{tabular*}{\linewidth}{@{\extracolsep{\fill}}cccc@{}}
\toprule
\textbf{Experience Level} & \textbf{Accuracy} & \textbf{Sensitivity} & \textbf{Specificity} \\
\midrule
Low experience (5 years) & 57.66\% & 58.24\% & 60.42\% \\
High experience (10 years) & 61.04\% & 61.50\% & 62.50\% \\
\bottomrule
\end{tabular*}
\end{table}

To evaluate the realism of our synthetic perirectal lymph nodes, we conducted a visual Turing test with two radiologists: one with five years of experience and the other with over ten years. Each radiologist independently assessed 385 samples (192 real and 193 synthetic), based on structural and textural features.

The low-experience radiologist achieved an \textbf{accuracy} of 57.66\%, while the high-experience radiologist attained 61.04\%. \textbf{Sensitivity}, which measures the ability to correctly identify synthetic nodes, ranged from 58.24\% to 61.50\%, while \textbf{specificity}, indicating the ability to correctly recognize real nodes, was between 60.42\% and 62.50\%. These results, only slightly above the 50\% expected from random guessing, suggest that synthetic and real nodes exhibit highly similar visual characteristics. Detailed performance metrics are provided in Table~\ref{tab:turing_test}.

\subsection{Visualization of Controllable Synthesis}
% !TEX root = ../top.tex
% !TEX spellcheck = en-US

\begin{figure}[tb]
        \centering
	\includegraphics[width=\linewidth]{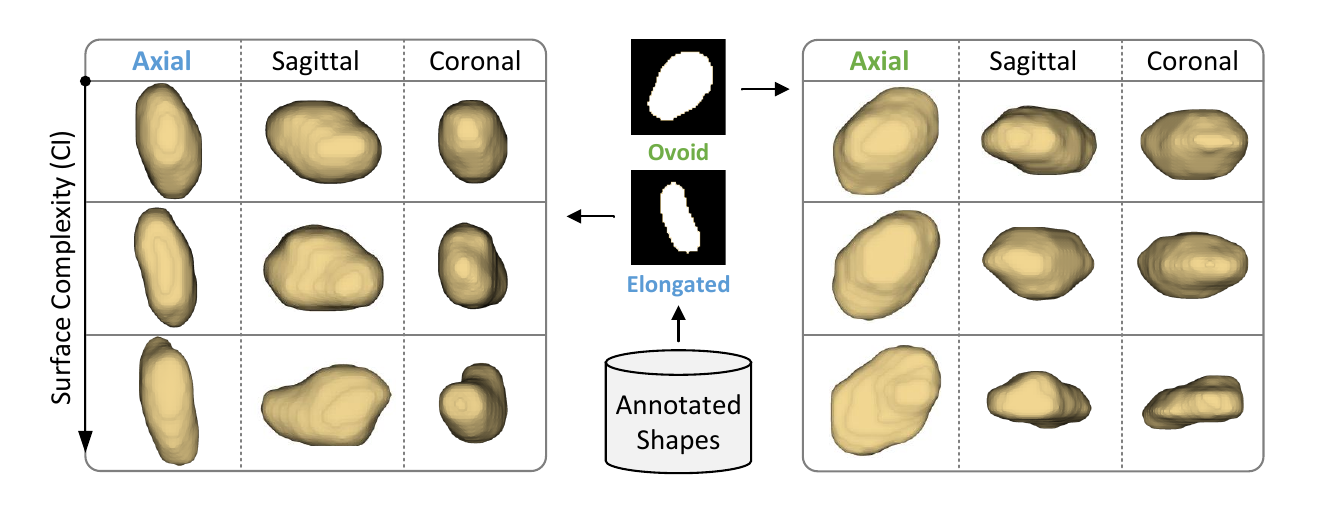}
        \caption{\textbf{Anatomical Guidance (AG).} This figure illustrates a framework for controlling 3D lymph node morphology, emphasizing both the global regulation of overall shape and the fine adjustment of surface details.}
	\label{fig:anatomical_guidance}
\end{figure}
\paragraph{\textbf{Anatomical Guidance (AG).}} AG allows for precise morphological control over perirectal lymph nodes by first selecting a shape template (e.g., elongated or ovoid) and then refining the surface complexity according to the Curvature Index (CI). Higher CI values produce more intricate surface details, enabling accurate, scalable shape adjustments, as illustrated in Figure~\ref{fig:anatomical_guidance}.

% !TEX root = ../top.tex
% !TEX spellcheck = en-US

\begin{figure}[tb]
        \centering
	\includegraphics[width=\linewidth]{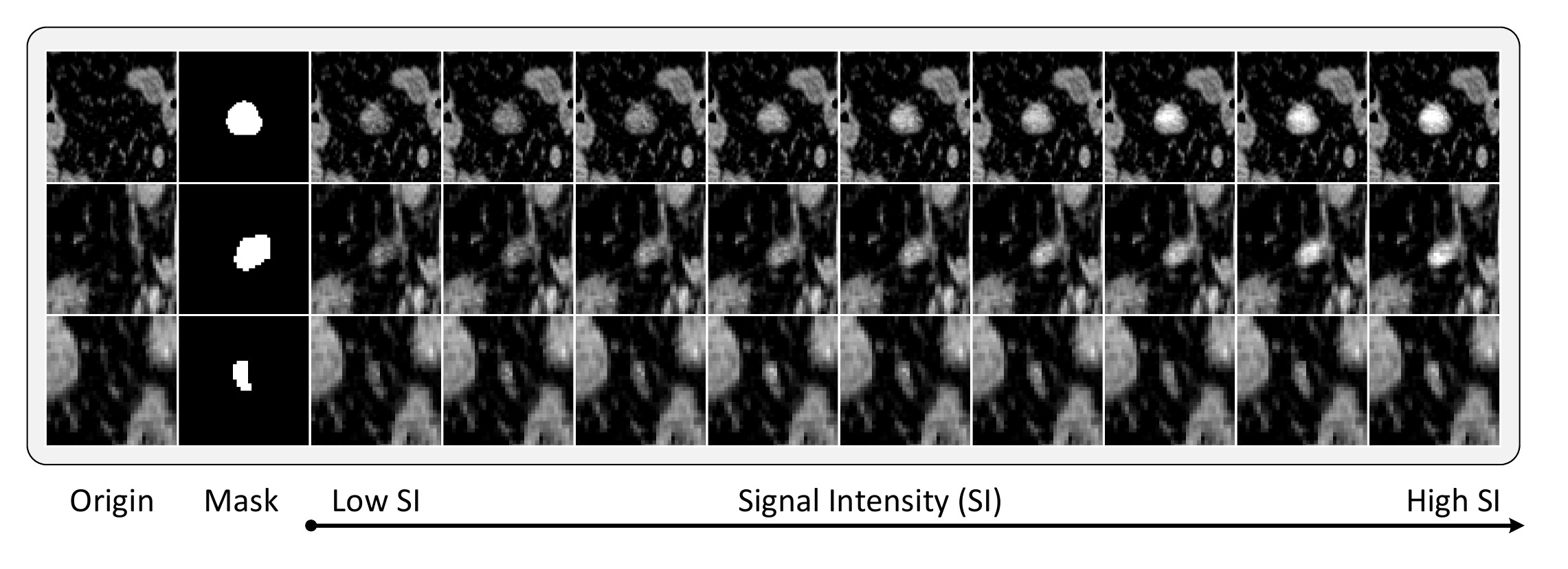}
    \caption{\textbf{Textural Guidance (TG).} Leveraging the synthetic mask, we controllably generate perirectal lymph nodes on real background images. The figure demonstrates lymph nodes with varying Signal Intensity (SI) values, from low to high, where higher SI values enhance the contrast against the background.}
    \label{fig:textural_guidance}
\end{figure}
\paragraph{\textbf{Textural Guidance (TG).}} By modulating Signal Intensity (SI) values, we control texture contrast levels for synthetic lymph nodes. This approach ensures realistic, diverse textures even for unseen masks, enabling robust control across different morphological variations, as shown in Figure~\ref{fig:textural_guidance}.
\section{Conclusion}

We propose \ourmodel{}, an SDF-based framework for the synthesis of perirectal lymph nodes, addressing the challenges of stability in morphological generation and controllability in the synthesis process. To enhance controllability, we introduce AG and TG, enabling precise modeling of complex lymph node structures and textures. \ourmodel{} effectively generates realistic and diverse lymph node structures with anatomically consistent masks and textures, providing a robust and controllable solution for perirectal lymph node synthesis and advancing the realism and utility of synthetic data in medical imaging.
% \bibliographystyle{splncs04}
% \bibliography{string,reference}

\printbibliography

@article{zhang2023care,
  title={CARE: A Large Scale CT Image Dataset and Clinical Applicable Benchmark Model for Rectal Cancer Segmentation},
  author={Zhang, Hantao and Guo, Weidong and Qiu, Chenyang and Wan, Shouhong and Zou, Bingbing and Wang, Wanqin and Jin, Peiquan},
  journal={arXiv preprint arXiv:2308.08283},
  year={2023}
}

@article{aljabri2022towards,
  title={Towards a better understanding of annotation tools for medical imaging: a survey},
  author={Aljabri, Manar and AlAmir, Manal and AlGhamdi, Manal and Abdel-Mottaleb, Mohamed and Collado-Mesa, Fernando},
  journal={Multimedia tools and applications},
  volume={81},
  number={18},
  pages={25877--25911},
  year={2022},
  publisher={Springer}
}

@article{li2018large,
  title={Large-scale retrieval for medical image analytics: A comprehensive review},
  author={Li, Zhongyu and Zhang, Xiaofan and M{\"u}ller, Henning and Zhang, Shaoting},
  journal={Medical image analysis},
  volume={43},
  pages={66--84},
  year={2018},
  publisher={Elsevier}
}

@article{perets2024inherent,
  title={Inherent Bias in Electronic Health Records: A Scoping Review of Sources of Bias},
  author={Perets, Oriel and Stagno, Emanuela and Yehuda, Eyal Ben and McNichol, Megan and Celi, Leo Anthony and Rappoport, Nadav and Dorotic, Matilda},
  journal={medRxiv},
  year={2024},
  publisher={Cold Spring Harbor Laboratory Preprints}
}

@article{zhang2023deep,
  title={Deep long-tailed learning: A survey},
  author={Zhang, Yifan and Kang, Bingyi and Hooi, Bryan and Yan, Shuicheng and Feng, Jiashi},
  journal={IEEE Transactions on Pattern Analysis and Machine Intelligence},
  volume={45},
  number={9},
  pages={10795--10816},
  year={2023},
  publisher={IEEE}
}

@article{wu2024medical,
  title={Medical long-tailed learning for imbalanced data: bibliometric analysis},
  author={Wu, Zheng and Guo, Kehua and Luo, Entao and Wang, Tian and Wang, Shoujin and Yang, Yi and Zhu, Xiangyuan and Ding, Rui},
  journal={Computer Methods and Programs in Biomedicine},
  pages={108106},
  year={2024},
  publisher={Elsevier}
}

@article{Savage2023SyntheticDC,
  title={Synthetic data could be better than real data},
  author={Neil Savage},
  journal={Nature},
  year={2023},
  url={https://doi.org/10.1038/d41586-023-01445-8}
}

@inproceedings{hu2023label,
  title={Label-free liver tumor segmentation},
  author={Hu, Qixin and Chen, Yixiong and Xiao, Junfei and Sun, Shuwen and Chen, Jieneng and Yuille, Alan L and Zhou, Zongwei},
  booktitle={Proceedings of the IEEE/CVF Conference on Computer Vision and Pattern Recognition},
  pages={7422--7432},
  year={2023}
}

@inproceedings{chen2024towards,
  title={Towards generalizable tumor synthesis},
  author={Chen, Qi and Chen, Xiaoxi and Song, Haorui and Xiong, Zhiwei and Yuille, Alan and Wei, Chen and Zhou, Zongwei},
  booktitle={Proceedings of the IEEE/CVF Conference on Computer Vision and Pattern Recognition},
  pages={11147--11158},
  year={2024}
}

@inproceedings{han2019synthesizing,
  title={Synthesizing diverse lung nodules wherever massively: 3D multi-conditional GAN-based CT image augmentation for object detection},
  author={Han, Changhee and Kitamura, Yoshiro and Kudo, Akira and Ichinose, Akimichi and Rundo, Leonardo and Furukawa, Yujiro and Umemoto, Kazuki and Li, Yuanzhong and Nakayama, Hideki},
  booktitle={2019 International Conference on 3D Vision (3DV)},
  pages={729--737},
  year={2019},
  organization={IEEE}
}

@article{jin2021free,
  title={Free-form tumor synthesis in computed tomography images via richer generative adversarial network},
  author={Jin, Qiangguo and Cui, Hui and Sun, Changming and Meng, Zhaopeng and Su, Ran},
  journal={Knowledge-Based Systems},
  volume={218},
  pages={106753},
  year={2021},
  publisher={Elsevier}
}

@article{wang2021realistic,
  title={Realistic lung nodule synthesis with multi-target co-guided adversarial mechanism},
  author={Wang, Qiuli and Zhang, Xiaohong and Zhang, Wei and Gao, Mingchen and Huang, Sheng and Wang, Jian and Zhang, Jiuquan and Yang, Dan and Liu, Chen},
  journal={IEEE Transactions on Medical Imaging},
  volume={40},
  number={9},
  pages={2343--2353},
  year={2021},
  publisher={IEEE}
}

@article{shin2018abnormal,
  title={Abnormal colon polyp image synthesis using conditional adversarial networks for improved detection performance},
  author={Shin, Younghak and Qadir, Hemin Ali and Balasingham, Ilangko},
  journal={IEEE Access},
  volume={6},
  pages={56007--56017},
  year={2018},
  publisher={IEEE}
}

@inproceedings{sharma2024controlpolypnet,
  title={ControlPolypNet: Towards Controlled Colon Polyp Synthesis for Improved Polyp Segmentation},
  author={Sharma, Vanshali and Kumar, Abhishek and Jha, Debesh and Bhuyan, MK and Das, Pradip K and Bagci, Ulas},
  booktitle={Proceedings of the IEEE/CVF Conference on Computer Vision and Pattern Recognition},
  pages={2325--2334},
  year={2024}
}

@article{abbood2022dr,
  title={DR-LL Gan: Diabetic Retinopathy Lesions Synthesis using Generative Adversarial Network.},
  author={Abbood, Saif Hameed and Abdull Hamed, Haza Nuzly and Mohd Rahim, Mohd Shafry and Alaidi, Abdul Hadi M and Salim ALRikabi, Haider TH},
  journal={International Journal of Online \& Biomedical Engineering},
  volume={18},
  number={3},
  year={2022}
}

@inproceedings{zhou2019high,
  title={High-resolution diabetic retinopathy image synthesis manipulated by grading and lesions},
  author={Zhou, Yi and He, Xiaodong and Cui, Shanshan and Zhu, Fan and Liu, Li and Shao, Ling},
  booktitle={International conference on medical image computing and computer-assisted intervention},
  pages={505--513},
  year={2019},
  organization={Springer}
}

@article{yao2021label,
  title={Label-free segmentation of COVID-19 lesions in lung CT},
  author={Yao, Qingsong and Xiao, Li and Liu, Peihang and Zhou, S Kevin},
  journal={IEEE transactions on medical imaging},
  volume={40},
  number={10},
  pages={2808--2819},
  year={2021},
  publisher={IEEE}
}

@article{zhu2004class,
  title={Class noise vs. attribute noise: A quantitative study},
  author={Zhu, Xingquan and Wu, Xindong},
  journal={Artificial intelligence review},
  volume={22},
  pages={177--210},
  year={2004},
  publisher={Springer}
}

@article{ju2022improving,
  title={Improving medical images classification with label noise using dual-uncertainty estimation},
  author={Ju, Lie and Wang, Xin and Wang, Lin and Mahapatra, Dwarikanath and Zhao, Xin and Zhou, Quan and Liu, Tongliang and Ge, Zongyuan},
  journal={IEEE transactions on medical imaging},
  volume={41},
  number={6},
  pages={1533--1546},
  year={2022},
  publisher={IEEE}
}

@article{kim2004high,
  title={High-resolution MR imaging for nodal staging in rectal cancer: are there any criteria in addition to the size?},
  author={Kim, Joo Hee and Beets, Geerard L and Kim, Myeong-Jin and Kessels, Alfons GH and Beets-Tan, Regina GH},
  journal={European journal of radiology},
  volume={52},
  number={1},
  pages={78--83},
  year={2004},
  publisher={Elsevier}
}

@inproceedings{curless1996volumetric,
  title={A volumetric method for building complex models from range images},
  author={Curless, Brian and Levoy, Marc},
  booktitle={Proceedings of the 23rd annual conference on Computer graphics and interactive techniques},
  pages={303--312},
  year={1996}
}

@article{ho2020denoising,
  title={Denoising diffusion probabilistic models},
  author={Ho, Jonathan and Jain, Ajay and Abbeel, Pieter},
  journal={Advances in neural information processing systems},
  volume={33},
  pages={6840--6851},
  year={2020}
}

@article{song2020denoising,
  title={Denoising diffusion implicit models},
  author={Song, Jiaming and Meng, Chenlin and Ermon, Stefano},
  journal={arXiv preprint arXiv:2010.02502},
  year={2020}
}

@inproceedings{park2019deepsdf,
  title={Deepsdf: Learning continuous signed distance functions for shape representation},
  author={Park, Jeong Joon and Florence, Peter and Straub, Julian and Newcombe, Richard and Lovegrove, Steven},
  booktitle={Proceedings of the IEEE/CVF conference on computer vision and pattern recognition},
  pages={165--174},
  year={2019}
}

@article{isensee2021nnu,
  title={nnU-Net: a self-configuring method for deep learning-based biomedical image segmentation},
  author={Isensee, Fabian and Jaeger, Paul F and Kohl, Simon AA and Petersen, Jens and Maier-Hein, Klaus H},
  journal={Nature methods},
  volume={18},
  number={2},
  pages={203--211},
  year={2021},
  publisher={Nature Publishing Group}
}

@inproceedings{ronneberger2015u,
  title={U-net: Convolutional networks for biomedical image segmentation},
  author={Ronneberger, Olaf and Fischer, Philipp and Brox, Thomas},
  booktitle={Medical image computing and computer-assisted intervention--MICCAI 2015: 18th international conference, Munich, Germany, October 5-9, 2015, proceedings, part III 18},
  pages={234--241},
  year={2015},
  organization={Springer}
}

@inproceedings{he2023swinunetr,
  title={Swinunetr-v2: Stronger swin transformers with stagewise convolutions for 3d medical image segmentation},
  author={He, Yufan and Nath, Vishwesh and Yang, Dong and Tang, Yucheng and Myronenko, Andriy and Xu, Daguang},
  booktitle={International Conference on Medical Image Computing and Computer-Assisted Intervention},
  pages={416--426},
  year={2023},
  organization={Springer}
}

@inproceedings{lugmayr2022repaint,
  title={Repaint: Inpainting using denoising diffusion probabilistic models},
  author={Lugmayr, Andreas and Danelljan, Martin and Romero, Andres and Yu, Fisher and Timofte, Radu and Van Gool, Luc},
  booktitle={Proceedings of the IEEE/CVF conference on computer vision and pattern recognition},
  pages={11461--11471},
  year={2022}
}

@inproceedings{achlioptas2018learning,
  title={Learning representations and generative models for 3d point clouds},
  author={Achlioptas, Panos and Diamanti, Olga and Mitliagkas, Ioannis and Guibas, Leonidas},
  booktitle={International conference on machine learning},
  pages={40--49},
  year={2018},
  organization={PMLR}
}

@article{chang2015shapenet,
  title={Shapenet: An information-rich 3d model repository},
  author={Chang, Angel X and Funkhouser, Thomas and Guibas, Leonidas and Hanrahan, Pat and Huang, Qixing and Li, Zimo and Savarese, Silvio and Savva, Manolis and Song, Shuran and Su, Hao and others},
  journal={arXiv preprint arXiv:1512.03012},
  year={2015}
}

@inproceedings{chou2023diffusion,
  title={Diffusion-sdf: Conditional generative modeling of signed distance functions},
  author={Chou, Gene and Bahat, Yuval and Heide, Felix},
  booktitle={Proceedings of the IEEE/CVF international conference on computer vision},
  pages={2262--2272},
  year={2023}
}

@article{bao2022conditional,
  title={Why are conditional generative models better than unconditional ones?},
  author={Bao, Fan and Li, Chongxuan and Sun, Jiacheng and Zhu, Jun},
  journal={arXiv preprint arXiv:2212.00362},
  year={2022}
}

@inproceedings{song2023loss,
  title={Loss-guided diffusion models for plug-and-play controllable generation},
  author={Song, Jiaming and Zhang, Qinsheng and Yin, Hongxu and Mardani, Morteza and Liu, Ming-Yu and Kautz, Jan and Chen, Yongxin and Vahdat, Arash},
  booktitle={International Conference on Machine Learning},
  pages={32483--32498},
  year={2023},
  organization={PMLR}
}

@STRING{ARXIV    = "arXiv Preprint"}

@STRING{NATURE   = "Nature"}
\end{document}